\documentclass[
superscriptaddress,
twocolumn,
nobibnotes,
altermagnetsmath,altermagnetssymb,
aps, 
prl,
showpacs,
letterpaper
]{revtex4-2}
\usepackage{xcolor}
\usepackage[colorlinks=true, linkcolor=purple, citecolor=purple, urlcolor=blue]{hyperref}
\usepackage{amsmath}
\usepackage{physics}
\makeatletter
\usepackage{graphicx}
\usepackage{dcolumn}
\usepackage{bm}

\makeatother

\newcommand{\sect}[1]{\vspace{0.3em}{\it #1.}---}

\begin{document}
\title{\textbf{Quantum Geometry of Altermagnetic Magnons Probed by Light} 
	}%
\author{Rundong Yuan}
\email{ry306@cam.ac.uk}
\affiliation{TCM Group, Cavendish Laboratory, University of Cambridge, J.\,J.\,Thomson Avenue, Cambridge CB3 0HE, United Kingdom}

\author{Wojciech J. Jankowski}
\email{wjj25@cam.ac.uk}
\affiliation{TCM Group, Cavendish Laboratory, University of Cambridge, J.\,J.\,Thomson Avenue, Cambridge CB3 0HE, United Kingdom}

\author{Ka Shen}
\affiliation{The Center for Advanced Quantum Studies, School of Physics and Astronomy, and Institute for Advanced Study, Beijing Normal University, Beijing 100875, China}
\affiliation{Key Laboratory of Multiscale Spin Physics, Ministry of Education, Beijing Normal University, Beijing 100875, China}

\author{Robert-Jan Slager}
\email{rjs269@cam.ac.uk}
\affiliation{Department of Physics and Astronomy, University of Manchester, Oxford Road, Manchester M13 9PL, United Kingdom}
\affiliation{TCM Group, Cavendish Laboratory, University of Cambridge, J.\,J.\,Thomson Avenue, Cambridge CB3 0HE, United Kingdom}

\newcommand{\TCM}{{TCM Group, Cavendish Laboratory, University of Cambridge, J.\,J.\,Thomson Avenue, Cambridge CB3 0HE, United Kingdom}}
\newcommand{\MCR}{Department of Physics and Astronomy, University of Manchester, Oxford Road, Manchester M13 9PL, United Kingdom}

\date{\today}
    
    \begin{abstract} 
    Magnons with momentum-dependent chirality are a key signature of altermagnets. We identify bicircular light as a smoking-gun optical probe for chiral altermagnetic magnons, selectively targeting their quantum geometry induced by an alternation of magnonic chirality.
    We show that in $d$-wave altermagnets, under a canting magnetic field, the altermagnetic magnons realize a nontrivial quantum geometry, resulting in an enhancement of the nonlinear second-order light-magnon interactions.
    We find that the scattering of bicircular pulses probes the present magnonic quantum geometry, even if the magnonic topology is trivial. 
    Hence, our findings establish bicircular Raman response as an optical effect of choice to identify altermagnetic magnons.
    As such, we propose a universal experimental protocol to distinguish altermagnets from antiferromagnets by detecting their magnon chirality patterns with light, independently of the underlying magnon topology.
    \end{abstract}

    \maketitle

    \sect{Introduction} Excitations in ordered media, such as magnets, provide insights into the physical character of the underlying order. 
    Magnons, the quanta of spin waves, carry spin-precessional information, such as chirality, in magnetic materials~\cite{kruglyak2010magnonics, chumak2015magnon, barman2021roadmap}.
    The dynamical chirality of magnons can be defined as the sum of the expectation value of right-handed (RH) and left-handed~(LH) spin-precessional angular momentum once the magnetic polarization is aligned by single-ion anisotropy or an external magnetic field~\cite{cheng2014spin}.
    Because of time-reversal symmetry~(TRS) breaking in ferromagnets, only RH magnons exist upon exciting a ferromagnetic ground state, resulting in chiral propagation of magnons~\cite{damon1961magnetostatic, chen2019excitation, yu2019chiral}
    and magnon spin currents with only one sign under a fixed magnetic field~\cite{ando2011inverse, maekawa2017spin, wang2024broad}.
    The LH magnon chirality is recovered in antiferromagnetic~(AFM) ground states, given their pseudo-TRS~\cite{cheng2014spin, li2020spin, wimmer2020observation, vaidya2020subterahertz}.

    In a recently proposed classification of magnetic materials~\mbox{\cite{Smejkal2022PRX, song2025altermagnets,vsmejkal2022emerging}}, altermagnets, the excitations of their magnetic order, i.e., altermagnetic (AM) magnons, possess chiral properties despite a vanishing net magnetic moment of the system, even without any chiral Dzyaloshinskii-Moriya interactions that split energies of RH and LH magnons in conventional antiferromagnets~\cite{vsmejkal2022emerging, vsmejkal2023chiral, liu2024chiral, sandratskii2025direct,jin2023cavity,jin2025strong, mook2014edge, mook2014magnon, cheng2014spin, cheng2016antiferromagnetic}.
    Unlike conventional antiferromagnets, the altermagnets break TRS by their spin space groups, and thus lift electronic band degeneracies according to their magnetic polarizations~\cite{vsmejkal2022emerging, vsmejkal2023chiral, liu2024chiral, sandratskii2025direct, reimers2024direct, ding2024large, krempasky2024altermagnetic, fedchenko2024observation}.
    Similarly, in altermagnets, the RH and LH magnons are energetically lifted on low-symmetry points of magnon band structures~\cite{vsmejkal2022emerging, vsmejkal2023chiral, liu2024chiral, sandratskii2025direct,jin2023cavity,jin2025strong}.
    This splitting of AM magnons, and their chirality, are induced by the asymmetric exchange interactions due to the alignment of nonmagnetic atoms between magnetic atoms~\cite{vsmejkal2022emerging, vsmejkal2023chiral, liu2024chiral,  sandratskii2025direct}.
    The chirality of magnetic excitations distinguishes altermagnets from antiferromagnets~\cite{vsmejkal2022emerging}, and thus gives insight into the classification of magnetic order on a lattice, which makes the detection of magnon chirality of AM magnon bands highly desirable~\cite{liu2024chiral, morano2025absence}.
    
    The chiral magnons in altermagnets have been investigated theoretically in various types of altermagnets such as $d$-wave altermagnet RuO$_2$~\cite{vsmejkal2023chiral}, Mn$_2$SeTe~\cite{yu2024chiral}, etc., and $g$-wave altermagnets $\alpha$-MnTe~\cite{sandratskii2025direct}, CrSb~\cite{zhang2025chiral}, $\alpha$-Fe$_2$O$_3$~\cite{hoyer2025altermagnetic}, etc.
    A~full classification of magnons can be dictated by the spin space group classifications~\cite{chen2025unconventional, Chenfangprx2024,xiao2024spin,chen2024enumeration,Schiff2025}, following similar strategies for (magnetic) space groups~\cite{Clas3, Clas4, Clas5, magnetic, magneticSIs,magnetictqc}.
    Experimentally, the energy splitting of magnons has been observed by inelastic neutron scattering in the $g$-wave altermagnet $\alpha$-MnTe and {$\alpha$-Fe$_2$O$_3$}~\cite{yu2024chiral,sun2025observation}. 
    Notably, the splitting of magnon bands was found to be absent in AM MnF$_2$~\cite{morano2025absence}.
    Electrical detection of magnon chirality has been conducted in \mbox{$\alpha$-Fe$_2$O$_3$}~\cite{sheng2025control}, but the excitation energy~($\sim 10$~GHz) is much lower than the splitting of chiral magnon bands, and thus makes their observations unlikely to originate from altermagnetism. 
    To our best knowledge, a direct probe of the split chirality at a certain frequency of AM magnons remains elusive until now.
    It should be stressed that the energy scale of chiral magnons is much higher than the range of electrical probes~\cite{wang2023long,el2023antiferromagnetic,sheng2025control, chen2025observation,chen2025deterministic,su2025enhancement}, whereas inelastic neutron scattering and conventional Brillouin light scattering experiments are not sensitive to AFM or AM magnon chirality~\cite{moussa1996spin, weber2019polarized, nambu2020observation,hamdi2023spin}.
  
   Centrally to this work, quantum geometry, consisting of the quantum metric and Berry curvature, can characterize the momentum-space Riemannian structure of Bloch wave functions arising from a given Hamiltonian~\cite{berry1989quantum, provost1980riemannian}.
    Nontrivial geometry is found to give rise to Hall conductances~\cite{niu1985quantized,  kane2005z, nagaosa2010anomalous, sinova2015spin, bouhon2023quantumgeometry,nakata2017magnonic, owerre2017noncollinear}, nonlinear transport~\cite{wang2023quantum, gao2023quantum}, and higher-order optical responses~\cite{schuler2020local, Topp2021, ahn2022riemannian, bouhon2023quantumgeometry, Bostrom2023, Kim2025} in various materials.
    In altermagnets, electronic quantum geometry has been reported to induce nontrivial topology~\cite{meng2024dirac,  syljuasen2025quantum, guo2025higher, lu2025signature}, nonlinear electron transport~\cite{fang2024quantum, han2025discovery}, magnetic instability~\cite{heinsdorf2025altermagnetic}, and magneto-optical signatures~\cite{chen2025magneto,li2025unconventional, li2026quantum}.
    On another note, bicircular light~(BCL), which targets information of different symmetry by its phase slip and frequency tunability, has been utilized to probe quantum geometries~\cite{PhysRevB.100.134301, Trevisan2022, ikeda2023photocurrent}.
    For magnons, the nontrivial topology of AFM spin waves could also be captured by light-matter coupling (LMC) under a canting~(perpendicular) magnetic field~\cite{Bostrom2023}.
    Although yet to be explored, LMC underpinned by the magnonic quantum metric, rather than the Berry curvature~\cite{Shen2020magnon,Liu2020dipolar,Bostrom2023, liu2023tunable, bermond2025}, promises to be pivotal for capturing  geometrical properties of magnons.
    Whether a~universal relation exists between magnon chirality distribution in altermagnets and magnonic quantum geometry is an open question, which we address here.

\begin{figure}[t!]
        \centering
        \includegraphics[width=\columnwidth]{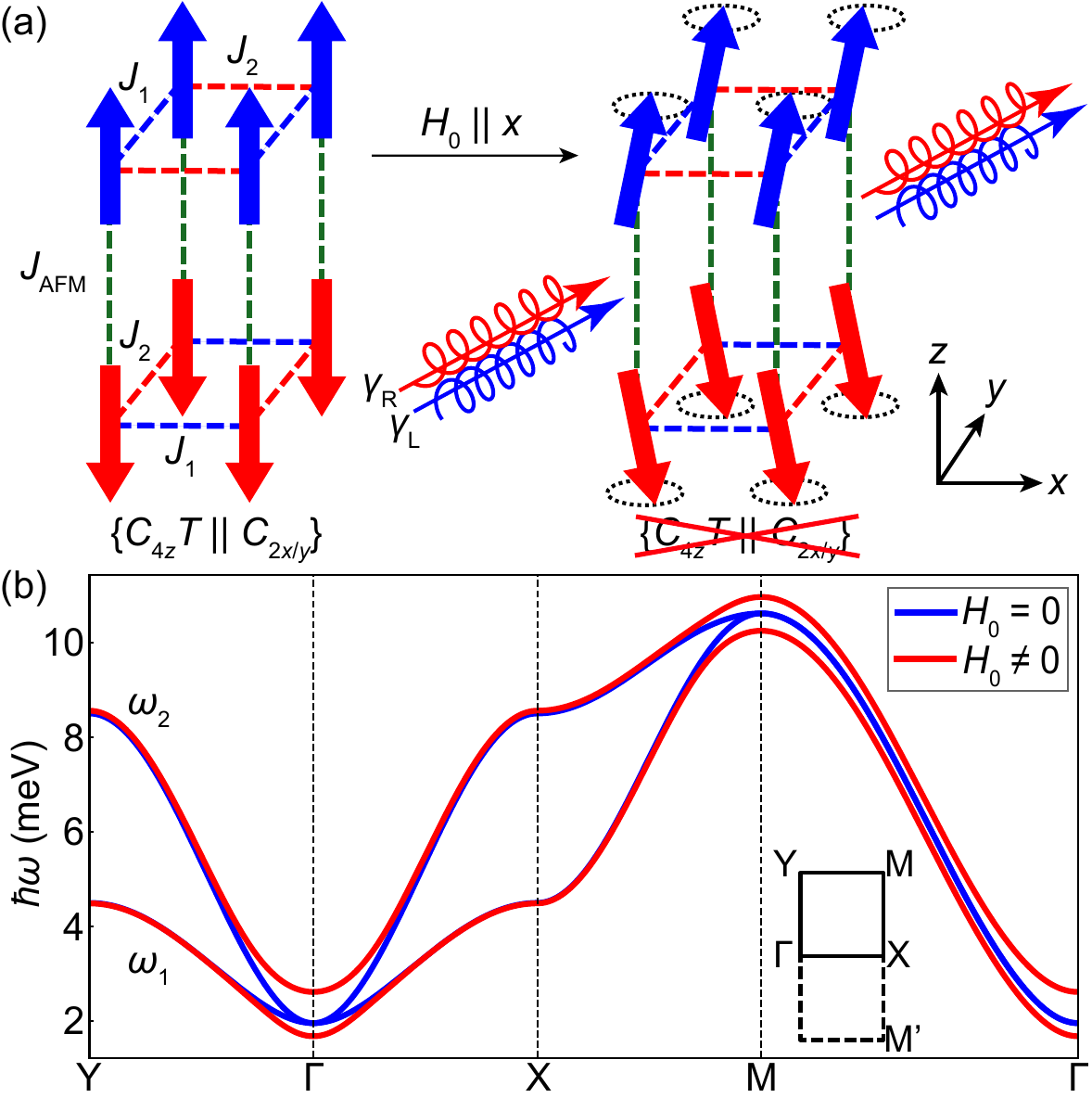}
        \caption{{\bf Modeling AM magnons with split chirality.} (a)~Minimal spin model for $d$-wave AM magnons, canted by an external magnetic field. The spin structures before (left) and after (right) applying a canting field. The AM magnons are bicircularly probed by photons of opposite handedness $(\gamma_\text{L/R})$. (b) Magnon band structure with~external canting field~(red lines) and without~({blue lines}). 
        Here, we set: $S = 1$, with $J_{1,2} = 1\pm0.5~\text{meV}$, $J_{\text{AFM}} = 2.0 ~\text{meV}$, $K = 0.8~\text{meV}$. $H_0 \neq 0$ to match the energy scale of $2.0~ \text{meV}$.
       The definitions of the parameters are further detailed in the End Matter.}
        \label{spin model}
    \end{figure}

    As shown in Fig.~\ref{spin model}(a), we propose an optical probe of chiral AM magnons and their quantum geometry with bicircular light~($\gamma_{\text{R}} +\gamma_{\text{L}}$) in an altermagnet canted by a magnetic field.
    We validate our findings in systems with trivial magnonic Chern numbers~\cite{Bostrom2023}, in a minimal $d$-wave AM Heisenberg model.
    After applying a canting magnetic field, a splitting of magnonic nodal line is found to appear at $\Gamma$-M high-symmetry line where the LH and RH magnons hybridize. 
    We show that in the first Brillouin zone (BZ), the magnon chirality switches as a function of  wave vector $\boldsymbol{k}$.
    As a consequence, the alternating chirality of AM magnons in the Bloch space gives rise to nontrivial geometry of magnon bands, alongside the presence of trivial band topology, which we reveal with accumulated phases in magnonic Wilson loop.
    We~deduce an enhancement of LMC through quantum metric in the dominant second and subleading fourth-order terms.
    Within our approach, the quantum geometry of AM magnons is revealed to be accessible  by nonlinear bicircular Raman spectroscopy.
    Such effects are found to be unique to AM systems, and are absent in AFM ones, as we further demonstrate.
    
   \sect{Modeling AM magnons} We start from a minimal \textit{d}-wave AM spin-lattice tight-binding model to investigate the AM magnons~\cite{jin2023cavity, jin2025strong}. 
   We consider an AM Heisenberg model canted by an external magnetic field without Dzyaloshinskii-Moriya interactions; see End Matter for technical details. 
   The model includes magnetic crystalline anisotropy to stabilize the magnetizations.
   As shown in the left panel of Fig.~\ref{spin model}(a), the considered spin Hamiltonian preserves a $C_{4z}T || C_{2x/y}$ symmetry without an external magnetic field, as expected for a \textit{d}-wave AM spin model, where $x/y$ denote $x$ or $y$ direction.
   Setting AM splitting $\Delta J = 0$, we can restore pseudo-TRS in the model and yield an AFM ground state.
   In the following, we will also show that the nontrivial geometrical effects are present regardless of the scale of exchange interactions, but only rely on the AM symmetry.
 
    We use a Holstein-Primakoff (HP) transformation for our AM model~\cite{holstein1940field}: $S_{x,i}^{\uparrow}+iS_{y,i}^{\uparrow}=\sqrt{2S}a_i$, $S_{x,i}^{\uparrow}-iS_{y,i}^{\uparrow}=\sqrt{2S}a^{\dagger}_i$, $S_{x,i}^{\downarrow}+iS_{y,i}^{\downarrow}=\sqrt{2S}b_i$, and $S_{x,i}^{\downarrow}-iS_{y,i}^{\downarrow}=\sqrt{2S}b^{\dagger}_i$, where  $a_i,b_i\left(a_i^{\dagger},b_i^{\dagger}\right)$ are magnon annihilation (creation) operators at the lattice sites $i$.
    Transforming the basis into Bloch space as ${a_i = 1/\sqrt{N} \sum_{\boldsymbol{k}} a_{\boldsymbol{k}}  e^{-i \boldsymbol{k}\cdot\boldsymbol{r}_i}}$ with position vectors $\boldsymbol{r}_i$, and analogously for the $b_i$, we obtain the quadratic bosonic Hamiltonian in the basis ${\psi_{\boldsymbol{k}}^{\dagger} = \left(a_{\boldsymbol{k}}^{\dagger} ,b_{\boldsymbol{k}}^{\dagger},a_{-\boldsymbol{k}},b_{-\boldsymbol{k}}\right)}$,
    with a linear-wave approximation. We only consider the in-plane wave vector $\boldsymbol{k} =(k_x, k_y)$ of magnons. The eigenmodes are obtained by a bosonic Bogoliubov--de Gennes~(BdG) transformation, $\psi_{\boldsymbol{k}} = U\phi_{\boldsymbol{k}}$, in which ${\phi_{\boldsymbol{k}}^{\dagger}=  \left(\alpha_{\boldsymbol{k}}^{\dagger} ,\beta_{\boldsymbol{k}}^{\dagger},\alpha_{-\boldsymbol{k}},\beta_{-\boldsymbol{k}}\right)}$ denotes the eigenmode basis.
    $U$ diagonalizes the Hamiltonian $\mathcal{H}(\boldsymbol{k})$ by ${U^{\dagger} \mathcal{H}(\boldsymbol{k}) U = \hbar\Omega(\boldsymbol{k})}$ with the magnonic eigenfrequency matrix $\Omega ({\boldsymbol{k}})=\text{diag}\left[\omega_1\left(\boldsymbol{k}\right),\omega_2\left(\boldsymbol{k}\right),\omega_1\left(-\boldsymbol{k}\right),\omega_2\left(-\boldsymbol{k}\right)\right]$~\cite{colpa1978diagonalization}, and $\omega_{i}\left(\boldsymbol{k}\right)$, with $i=1,2$, denote the energies of magnons with a wave vector $\boldsymbol{k}$. 
    The transformation matrix $U$ is normalized by $ \bar{U}U^{\dagger}  = I$, where ${\bar{U}^\dagger \equiv \tau_3 U^\dagger \tau_3}$ is the paraunitary conjugate of $U$ in BdG transformation \mbox{(see End Matter)}.
    The model supports ${\omega_i= \omega_{\bar{i}}}$, where $\bar{i}$ denotes the index with a opposite wave vector, as it obeys the $C_{2z}$ symmetry, which culminates in $\left(C_{4z}T\right)^2 = 1$. 
    This indicates a trivial topology of considered magnon bands, since the symmetry obstructs the existence of nontrivial Pfaffian operators at high-symmetry points~\cite{araya2023pfaffian}.
    Moreover, our model can be adiabatically transformed into a nonrelativistic AFM model, while preserving the $C_{4z}T$ symmetry, meaning that both belong to the same topological class.
    It has been investigated that the nonrelativistic AFM magnons have trivial topology~\cite{mook2014edge, nakata2017magnonic, owerre2017noncollinear}, which is consistent with the model studied hereby, which realizes topologically trivial magnons.
    
    In Fig.~\ref{spin model}(b), we show the magnon band structure of the $d$-wave AM model with and without the perpendicular magnetic field (red and blue lines).
    The high-symmetry points are defined in the right-lower panel of Fig.~\ref{spin model}(b).
    Consistent with previous findings~\cite{jin2023cavity,jin2025strong, vsmejkal2023chiral}, we find split chirality for magnon bands.
    Without an external magnetic field, a magnonic nodal line appears at the $\Gamma$-M high-symmetry line, and a Dirac-type band touching appears at the $\Gamma$ point.
    The perpendicular field lifts the degeneracy at the nodal line and at the $\Gamma$ point, which aligns with our expectation, since the field breaks the $C_{4z}T||C_2$ symmetry, as shown in Fig.~\ref{spin model}.
    As will be shown below, the canting field is essential for the identification of magnonic quantum geometry.

\begin{figure}[t!]
        \centering
        \includegraphics[width=0.49\textwidth]{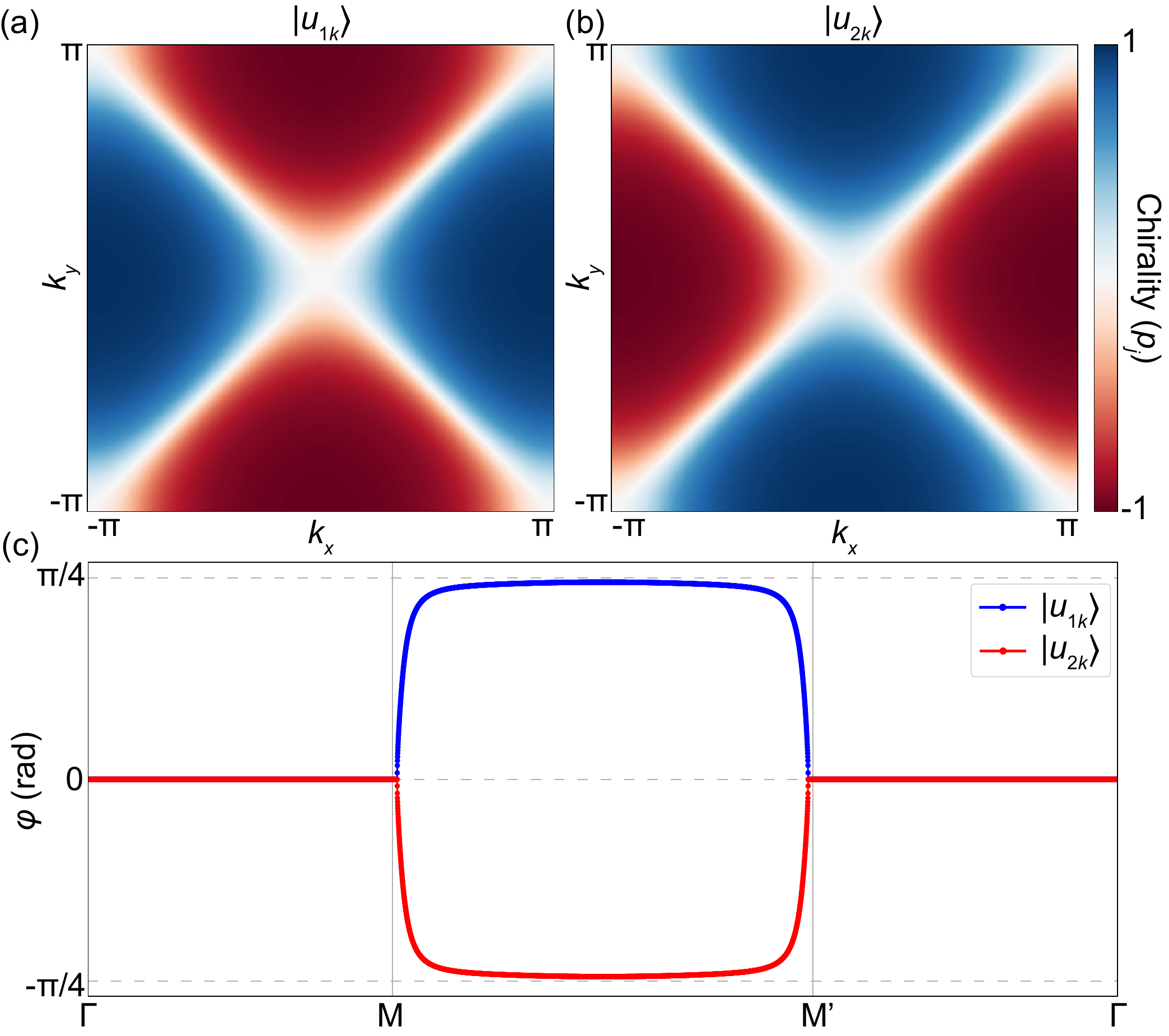}
        \caption{{\bf Interplay of magnonic chirality and quantum geometry in altermagnets.} (a),(b) The chirality of lower and upper magnon bands $\ket{u_{1\boldsymbol{k}}}  \equiv [U(\boldsymbol{k})]_1$, $\ket{u_{2\boldsymbol{k}}} \equiv [U(\boldsymbol{k})]_2$. The magnon chirality over momentum space $(p_j)$ is presented, with the blue and red colors representing the RH and LH chirality, respectively. (c) Magnonic Wilson lines along IBZ. The blue and red curves denote Wilson lines for $\ket{u_{1\boldsymbol{k}}}$ and $\ket{u_{2\boldsymbol{k}}}$. The parameters are the same as in Fig.~\ref{spin model}.}
        \label{geometric}
    \end{figure}

    \sect{Chirality and quantum geometry of AM magnons} Since the split AM magnon bands are chiral~\cite{vsmejkal2023chiral}, it is important to quantify their chirality $p_j$, which is shown in Figs.~\ref{geometric}(a),(b).
    The chirality of AM magnons can be inferred from their angular momentum ${\left\langle S^z \right\rangle_j =   \left\langle S_a^z \right\rangle_j - \left\langle S_b^z \right\rangle_j}$, where $\langle\ldots\rangle_j$ denotes the expectation value associated with magnon band index $j$. Subscripts $a/b$ denote the spin precession of sublattice $a^{(\dagger)}$/$b^{(\dagger)}$, which provide constant and definite (opposite) angular momenta.
    Since the equilibrium magnetization vanishes, one can further define the normalized chirality upon HP and BdG transformations, 
    \begin{equation}
        p_j\equiv \sum_{i,\text{RH}}\left|U^{-1}_{ij}\right|^2 - \sum_{i,\text{LH}}\left|U^{-1}_{ij}\right|^2,
    \end{equation} which represents the difference between the amplitude of RH and LH magnons~\cite{vsmejkal2023chiral}.
    Here, $p_j > 0~(p_j<0)$ represents an RH (LH) magnon chirality, and $p_j = 0$ represents a linear polarization.
    On the high-symmetry line $\Gamma$-M, one can observe that although the magnon bands are lifted, the chirality of both bands is similarly linear in our model.
    What~is more, the distribution of chirality $p_j$ over the first BZ behaves distinctly in AM and AFM; see Supplemental Material (SM) for details~\cite{SM}.

    We emphasize that the LH and RH characters of magnons are directly captured by the amplitudes in the original basis ($a_{\boldsymbol{k}}/b_{\boldsymbol{k}}$) of the system, i.e., in the column of the BdG matrix that denotes the eigenvectors of magnons in the Bloch space.
    Meanwhile, given the alternating feature of $p_j$ in momentum space, it is worth inspecting quantum-geometric consequences of the distribution of magnon chirality, since the quantum geometry relies on the differential features of the eigenvectors~\cite{provost1980riemannian,berry1989quantum}. 
    To characterize the geometry and topology of the modeled AM magnons, we calculate the gauge-invariant magnonic Wilson loop [Fig.~\ref{geometric}(c)] along a closed path surrounding the irreducible Brillouin zone~(IBZ), whose quantization is guaranteed by the $C_{4z}T$ symmetry in our model~\cite{araya2023pfaffian, alex2014wilson, bouhon2019wilson}.
    The Wilson loop is deduced from the phases of the Wilson line: ${W\left(\boldsymbol{k}\right) = \prod_{ \partial \text{IBZ}} \bar{U}^\dagger\left(\boldsymbol{k}_{i+1}\right) U\left(\boldsymbol{k}_{i}\right)}$,
    where $\boldsymbol{k}_i$ denote the wave vectors at different points of the Wilson loop contour.
    Importantly, we use a different definition from fermionic Wilson loop, since the overlap of neighbor bosonic BdG eigenvector should be represented by $ \bar{U}\left(\boldsymbol{k}_{1}\right) U\left(\boldsymbol{k}_{2}\right)$.
    The magnonic Wilson loop eigenvalue moves from $\phi=0$ to $\phi=\pm\pi/4$, and back to $\phi = 0$, which indicates the eigenvectors of the magnon bands accumulate a Berry phase, which is subsequently dissolved around the loop over the boundary of IBZ; see Fig.~\ref{geometric}(c). 
    The phase accumulated along the Wilson loop is also found to be parameter-independent (see SM~\cite{SM}), indicating that the nontrivial quantum geometry is protected by the model symmetry~\cite{araya2023pfaffian, alex2014wilson, bouhon2019wilson}, rather than arising from specific parameter choices.
    The magnonic Wilson loops are not winding around the IBZ, in agreement with our previous discussion of topology.
    Meanwhile, in an AFM model, the Wilson loop does not possess any nontrivial accumulated phases~(see SM~\cite{SM}).

    \begin{figure}[t]
    \centering
    \includegraphics[width=\columnwidth]{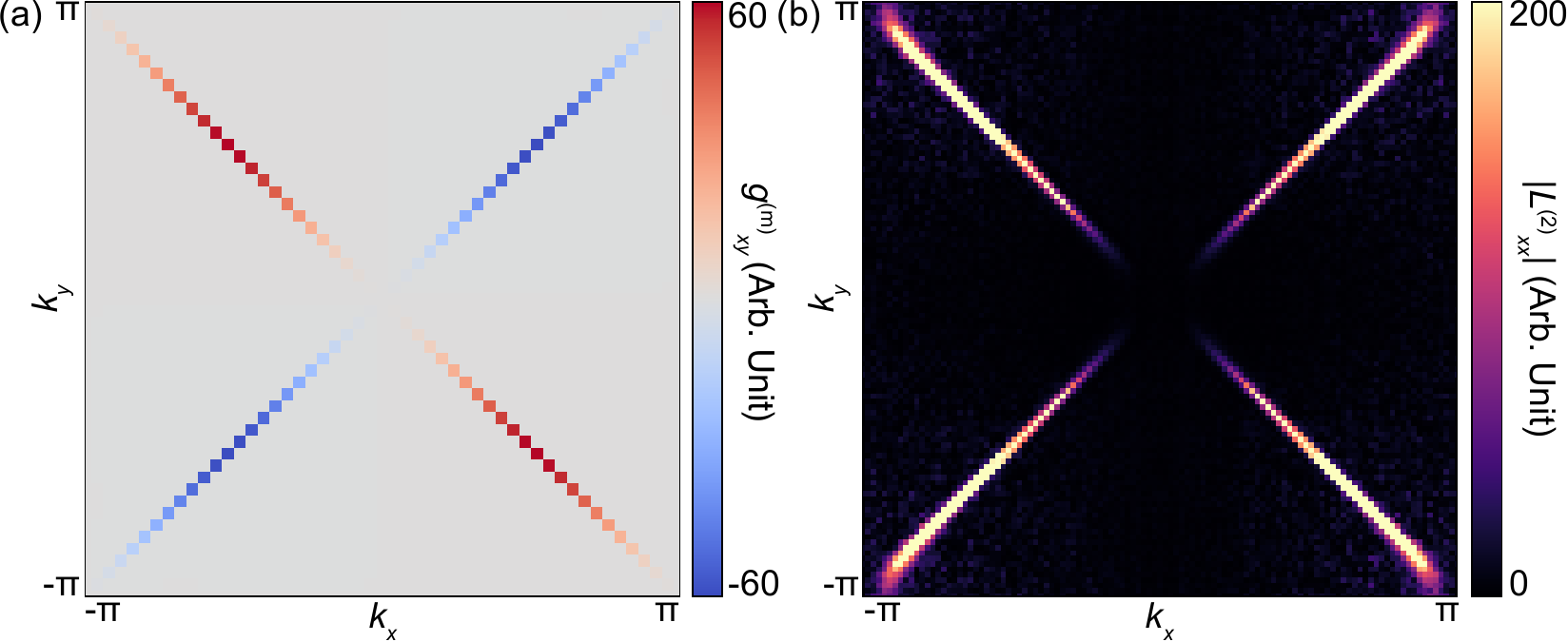}
        \caption{{\bf Quantum metric and nonlinear light-magnon coupling in a \mbox{$d$-wave} altermagnet.} (a) The quantum metric of magnons $g^{(\text{m})}_{ab}$ in the lower band of the canted minimal $d$-wave AM model, over the first Brillouin zone. The component of magnonic metric $g^{(\text{m})}_{xy}$ peaks at the chirality ($p_j$) crossover boundary. (b) The momentum-space features of the second-order LMC term $L^{(2)}_{xx}$. $L^{(2)}_{ab}$, which is enhanced by the magnonic quantum metric, is maximal at the magnon chirality crossover regions. The parameters are the same as in Fig.~\ref{spin model}, except that $H_0$ matches the energy of $0.5$~meV. }
        \label{LMC}
    \end{figure}

    We confirm that the AM magnons with trivial Chern number carry nontrivial quantum geometry, by an explicit evaluation of the quantum metric tensor; see Fig.~\ref{LMC}(a). 
    The quantum geometry of magnons is captured by magnonic quantum geometric tensor, ${Q^{(\text{m})}_{ab} = g^{(\text{m})}_{ab} - \frac{\text{i}}{2} \Omega^{(\text{m})}_{ab}}$~\cite{Bostrom2023}. 
    Here, we use the superscript (m) to denote the magnonic nature of the quantum geometric tensor considered.
    In particular, the magnonic Berry curvature $\Omega^{(\text{m})}_{ab}$, which naturally presents in topologically nontrivial magnons such as Chern magnons, can be characterized by dichroic Raman scattering~\cite{Bostrom2023}. 
    For topologically trivial magnons, the presence of any Berry curvature is not underpinned by Chern topology. 
    Here, the characteristic signature arising from the unique alternation of chirality of AM magnons across BZ~\cite{vsmejkal2023chiral} is the nontrivial magnonic quantum metric $g^{(\text{m})}_{ab}$.  
    The enhancement of the magnonic quantum metric is guaranteed by the lower bound associated with the derivative of  magnon chirality, as described in the End Matter.
    The canting field that split the magnon nodal line is found to be essential here given the definition of the magnonic metric~[see End Matter, Eq.~\eqref{metric}].
    Given that the quantum metric $g^{(\text{m})}_{ab}$ can also be probed by the LMC~\cite{gao2023quantum,Kim2025}, we focus on the metric-induced optical \mbox{features} associated with the nonlinear magnonic LMC in the following.

    \sect{Light-magnon interaction in altermagnets}
    We now construct a methodology to probe the quantum geometry in AM magnons with light.
    At optical frequencies, the magnonic LMC can be described by the Fleury-Loudon (FL) mechanism~\cite{fleury1968scattering,Bostrom2023}, where the photon electric field couples to spin-dependent electric dipoles.
    \mbox{Although} magnons do not naturally admit a direct Peierls substitution~\cite{eckstein2017designing,li2018theory}, the FL term generates an effective coupling,
    $\mathcal{H}(\boldsymbol{k},\boldsymbol{A})=\mathcal{H}(\boldsymbol{k}-\boldsymbol{A})$ (see SM~\cite{SM}), with ${\boldsymbol{A}=(A_x,A_y)}$ the vector potential of light.
    Accordingly, under the long-wavelength approximation, i.e., $
   \boldsymbol{A}\cdot\boldsymbol{d}_{ij}\ll 1$, with $\boldsymbol{d}_{ij}$ denoting the position vector from spin $i$ to $j$, the light-magnon coupling can be expanded as~\cite{fleury1968scattering,Bostrom2023}:
    $
    {\mathcal{H}(\boldsymbol{k},\boldsymbol{A})=
    \mathcal{H}(\boldsymbol{k})
    +
    L^{(1)}_{a} A_a
    +
    L^{(2)}_{ab} A_a A_b
    +
    \ldots,}$ and the second-order LMC Hamiltonian is given by 
        \begin{equation}\label{LMC2}
            \begin{aligned}
                L_{ab}^{\left(2\right)} =& -\Omega g_{ab}^{(\text{m})}+U_a^\dagger \bar{U}  \Omega  \bar{U}^\dagger U_b 
                - \Omega _a \bar{U}^\dagger U_b + U^{\dagger} \bar{U}_a \Omega _b  
               \\ 
                &+\frac{1}{2} \left(U^{\dagger} \bar{U}_{ab} \Omega  + \Omega _{ab} -\Omega  \bar{U}^{\dagger}U_{ab} \right)+\left(a \leftrightarrow b \right),
            \end{aligned}
    \end{equation}where the subscripts $a,b$ in the right-hand side give ${U_a \equiv \partial_{k_a} U}$, ${U_{ab} \equiv \partial_{k_a} \partial_{k_b}U}$, and analogously $\Omega_{a}\equiv \partial_{k_a} \Omega$, $\Omega_{ab}\equiv \partial_{k_a} \partial_{k_b} \Omega$.
    The further contributions are negligibly small, whereas are also calculated up to the fourth order for completeness; see SM~\cite{SM}.
    One finds that the quantum metric enters in the first term of $L_{ab}^{(2)}$, thus confirming that the emergent quantum metric in the magnon bands can enhance the nonlinear LMC~\cite{Bostrom2023}. 
    In Fig.~\ref{LMC}, we show the distribution of quantum metric and LMC strength as a function of wave vector around the first BZ.
    In Fig.~\ref{LMC}(a), the metric is found maximal at the lifted $\Gamma$-M line. 
    The absolute value of~$L_{xx}^{\left(2\right)}$ in the perturbative Hamiltonian is also calculated for the first diagonal term, as shown in Fig.~\ref{LMC}(b).
    Although Eq.~\eqref{LMC2} contains additional terms beyond the quantum metric, our calculations reveal a direct correspondence between the maxima of the quantum metric and the LMC strength, confirming the interplay between nontrivial quantum geometry and nonlinear Raman response of AM magnons.
    Furthermore, in a $PT$-symmetric AFM model with trivial magnon topology, where the alternation of magnon RH and LH chirality is prohibited, a simultaneous vanishing of the quantum metric and the LMC terms is observed (see SM~\cite{SM}).
    We also note that the canting field is introduced only to lift the degeneracy on $\Gamma$-M high-symmetry line, rather than to generate the LMC terms. 
    Indeed, the LMC strength decreases with an increasing canting field (see SM~\cite{SM}).
    We then establish the BCL scattering as an experimental probe for the LMC in altermagnets, given its tunability on the polarizations of the incident light, which allows to target~$L_{ab}^{\left(2\right)}$.

    \begin{figure}[t!]
        \centering
        \includegraphics[width=\columnwidth]{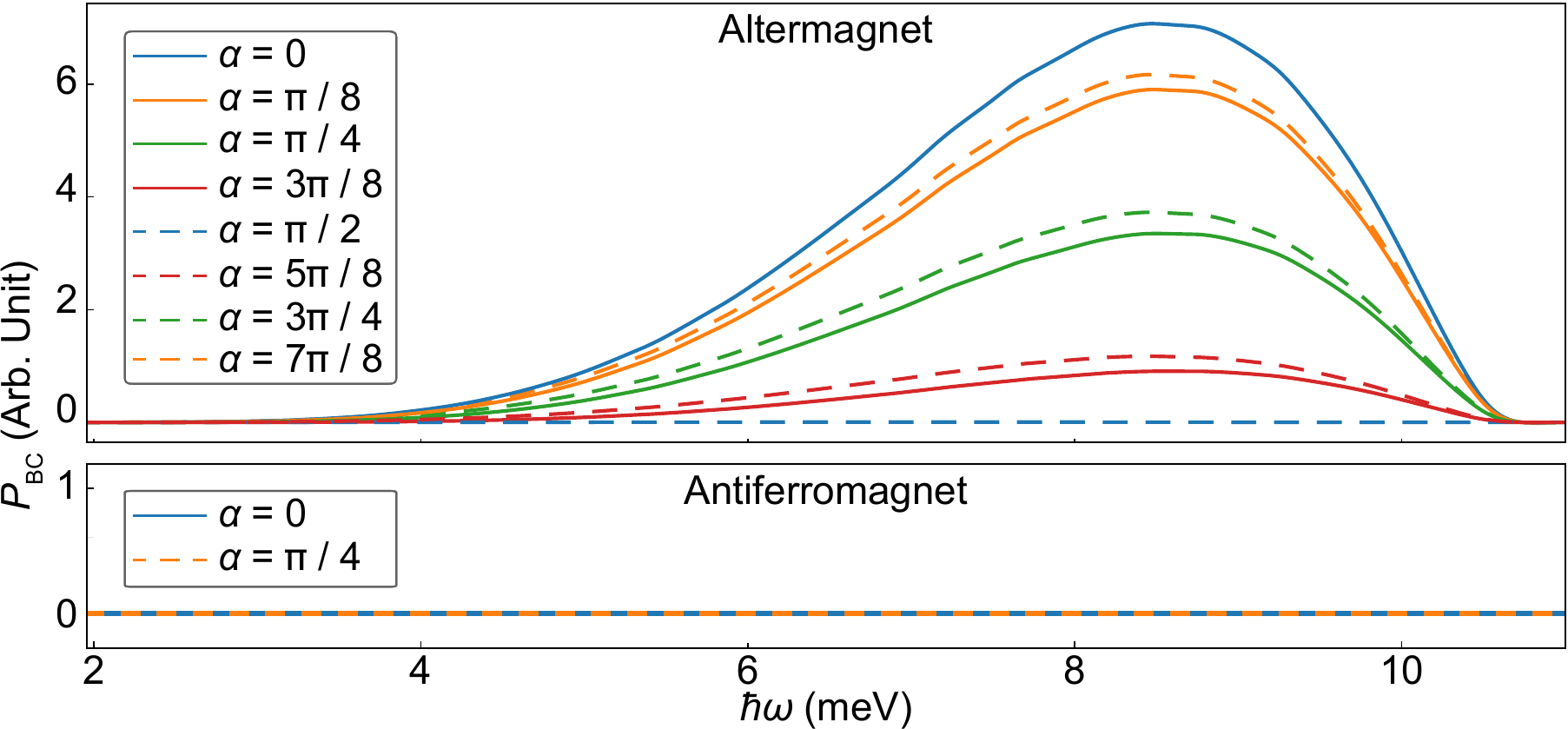}
        \caption{{\bf Cross section of magnonic LMC with bicircular light.} Bicircular light of frequency $\omega_\text{in}$ with phase shift $\alpha$ between left- and right-circular components is inelastically scattered to frequency $\omega_\text{sc}$, experiencing a Raman shift $\omega = \omega_\text{in} - \omega_\text{sc}$. Resolution of the light-AM magnon cross~section $P_\text{BC}$ capturing the scattering of AM magnons from light against $\alpha$ and $\omega$ allows to retrieve $L^{(2)}_{ab}$ enhanced by the magnonic quantum metric $g^{(\text{m})}_{ab}$ reflecting the chirality pattern. The parameters are the same as in Fig.~\ref{LMC}.}
        \label{BC}
    \end{figure}

\sect{Probing AM magnons with bicircular light} 
    We now show how BCL can be used to probe the nontrivial magnonic geometry of AM magnons within realistic experimental setups. 
    The BCL field takes the form ${\boldsymbol{A}(t) = \boldsymbol{A}_\text{L} e^{\text{i} (\eta \omega_\text{in} t + \alpha)} + \boldsymbol{A}_\text{R} e^{-\text{i} \omega_\text{in} t}}$~\cite{PhysRevB.100.134301,Trevisan2022}, including left and right circularly polarized light components with frequencies~$\omega_\text{in}$. 
    For ${\alpha = 0}$, the nonlinear LMC of AM magnons to linearly polarized light is retrieved from the bicircular coupling.
    The combinations of left and right circularly polarized beams, yielding linearly polarized light, have been previously proven successful in probing nontrivial electronic quantum geometry~\cite{Jankowski2025}.
    The nonlinear Raman response is estimated by ${P_{\text{BC}}\propto\sum_i\left| (\bar{U}H_{\text{R}}U)_{i\bar{i}}\right|^2\delta\left(\omega_i- \omega_{\text{sc}}+\omega_{\text{in}}\right)}$, where $\omega_{\text{in}}$ and $\omega_{\text{sc}}$ are the incident and scattered light frequencies, respectively, and $H_{\text{R}}$ is the Raman scattering Hamiltonian given by ${H_{\text{R}}\propto\sum_{ij}(\boldsymbol{A}^{*}_{\text{sc}}\cdot\boldsymbol{d}_{ij}})(\boldsymbol{A}_{\text{in}}\cdot\boldsymbol{d}_{ij})\mathcal{H}(\boldsymbol{k})$ (see~also~SM~\cite{SM}). 
    To propose a credible realization for experiments, we set $\eta=1$ and assume that the incident and scattered photons have the same polarizations.  
    We can reconstruct $L^{(2)}_{ab}$ on filtering out the first-order contribution $L^{(1)}_a$, which is distinctly sensitive to phase delay $\alpha$ of BCL (see SM~\cite{SM}). In Fig.~\ref{BC}, we show how the magnonic LMC cross~section $P_{\text{BC}}$~\cite{Bostrom2023} measured as a~function of $\alpha$ allows to probe the magnonic metric $g^{(\text{m})}_{ab}$, which captures the momentum-space chirality crossovers unique to the AM magnons. 
    Transforming LMC terms to a circular basis~(see SM~\cite{SM}), one finds that the dependence of cross~section $P_{\text{BC}}$ on phase \mbox{difference} $\alpha$ consistently captures the interplay between the Raman \mbox{response} and the polarizations of the incident and scattered light, which reflects the AM magnon chirality. 
    A~single-beam \mbox{linearly} polarized laser, having an equal superposition of left  and right circularly polarized components, can in principle also detect such an AM LMC signal.
    However, it does not permit tracking the signal as a function of the relative phase introduced in our proposal.
    \mbox{Notably}, we also observe that the AFM phase lacks similar \mbox{features} (see Fig.~\ref{BC}), consistent with degenerate LH and RH \mbox{chirality}~(see SM~\cite{SM}).
    Physically, our formalism probes the two-magnon processes via Raman scattering, which is insensitive to the wave vector of magnons, and thus cannot resolve detailed magnon dispersions, unlike inelastic neutron scattering measurements~\cite{fleury1968scattering,Brockhouse1957scattering}.
    
    \sect{Discussion and conclusions} We discuss the applicability of our results. 
    On the material side, the LMC effects of AM magnons could be realized in various types of AM materials, given that our proposed experimental signatures do not rely on a specific lattice. 
    In materials with strong spin-orbit coupling, non-FL contributions can dominate the magnonic Raman response~\cite{Yang2021non}, and the candidate materials should therefore be chosen to avoid such regime. 
    Suitable examples include the $d$-wave altermagnets RuO$_2$~\cite{vsmejkal2023chiral} and MnF$_2$~\cite{morano2025absence}, as~well~as LuFeO$_3$~\cite{galindez2025altermagnetic,yang2026altermagnet}, and the \mbox{$g$-wave} altermagnet $\alpha$-Fe$_2$O$_3$~\cite{hoyer2025altermagnetic,sun2025observation}.
    Their magnon energy scale is typically around $0$--$100$~meV, as reported in the above \mbox{references}.
    Depending on the different lattice symmetries, e.g. $g$- $f$-, $i$-wave, we expect analogous enhancements of LMCs with different symmetries over $\boldsymbol{k}$~space to be manifested in the scattering of BCL.
    We note that other types of AM models, such as honeycomb and checkerboard altermagnets~\cite{meng2024dirac, syljuasen2025quantum, guo2025higher}, can also accumulate nontrivial magnonic Berry phases and realize topological magnons.
    We expect that such nontrivial quantum geometry can likewise be observed using BCL. 
    Additionally, magnon-magnon interactions in altermagnets are expected to renormalize the magnon bands and induce chirality-dependent anisotropic damping, which may provide additional signatures of the AM order~\cite{antonio2025giant, Garcia2025magnon, jin2026interaction}. 
    Extending the magnonic LMC theories beyond linear spin-wave theory, e.g., including three-magnon processes, remains a~promising direction for future works. Our~results show that the LMC effects can clearly distinguish altermagnets from conventional antiferromagnets due to a qualitative difference between the AM and AFM magnons, as reflected by their distinct quantum \mbox{geometries}. 
    Our findings provide an optical criterion for confirming the presence of AM order with its unique spin excitations. 

\sect{Note added} We recently became aware of Ref.~\cite{liu2026light}, where Liu,~\textit{et~al.}~use topological ferromagnetic magnons to showcase probing magnonic quantum geometry with~LMC.

    \sect{Acknowledgments} We thank Bo Peng, Habib Rostami, Haiming Yu, and Yuelin Zhang for fruitful discussions. 
    We further thank Ting Qiu for spotting typos. 
    R.Y.~acknowledges the support by CSC Scholarship No. 202408060249 and Cambridge Commonwealth, European, and International Trust. 
    W.J.J.~acknowledges funding from the Rod Smallwood Studentship at Trinity College, Cambridge. 
    K.S. acknowledges support from the National Natural Science Foundation of China (Grants No. 12374100, and No. 12550004) and the \mbox{Fundamental} Research Funds for the Central Universities. R.-J.S. \mbox{acknowledges} funding from an EPSRC ERC underwrite grant  EP/X025829/1 and a Royal Society exchange grant IES/R1/221060 as well as Trinity College, Cambridge.
    
\appendix

\bibliography{references}

\section*{End Matter}

\sect{Appendix A: Details of the AM spin Hamiltonians}  Below we provide further technical details on the \mbox{$d$-wave} altermagnet model studied in this work~\cite{jin2023cavity}. In real space, the spin Heisenberg Hamiltonian $\mathcal{H}$ written in terms of the lattice spin operators $\boldsymbol{S}_i^{\uparrow},\boldsymbol{S}_i^{\downarrow}$, which we consider in this work, reads,
    \begin{equation}
    \begin{aligned}
    \mathcal{H} = &-\left(\sum_{\left\langle ij\right\rangle_x}J_1 \boldsymbol{S}_i^{\uparrow} \cdot \boldsymbol{S}_j^{\uparrow} + \sum_{\left\langle ij\right\rangle_y}J_2 \boldsymbol{S}_i^{\uparrow} \cdot \boldsymbol{S}_j^{\uparrow}\right)
   \\&-\left(\sum_{\left\langle ij\right\rangle_x}J_2 \boldsymbol{S}_i^{\downarrow} \cdot \boldsymbol{S}_j^{\downarrow} + \sum_{\left\langle ij\right\rangle_y}J_1 \boldsymbol{S}_i^{\downarrow} \cdot \boldsymbol{S}_j^{\downarrow}\right) 
    \\&-\sum_i J_{\text{AFM}}\boldsymbol{S}_i^{\uparrow} \cdot \boldsymbol{S}_i^{\downarrow} \\ 
    &-\frac{1}{2}K\sum_i \left[ (\boldsymbol{S}_i^{\uparrow}\cdot \hat{\boldsymbol{z}})^2 + (\boldsymbol{S}_i^{\downarrow}\cdot \hat{\boldsymbol{z}})^2\right] 
		\\&-g \mu_{\text{B}}H_0 \sum_i\left(\boldsymbol{S}_i^{\uparrow}+\boldsymbol{S}_i^{\downarrow} \right)\cdot \hat{\boldsymbol{x}}, 
    \end{aligned}
\end{equation}  
with $J_{1,2} = J \pm \Delta J$ the intrasublattice exchange coupling coefficient, $J_{\text{AFM}} < 0$ the intersublattice AFM exchange coupling coefficient, $K>0$ the easy-axis magnetic anisotropy, $\hat{\boldsymbol{x}}, \hat{\boldsymbol{z}}$ the unit vectors in the $x$ and $z$ directions, the Bohr magneton $\mu_B$, Land\'e $g$~factor, and $H_0$ an external field along $x$ direction, which cants the sublattices; see Fig.~\ref{spin model}(a). $\Delta J$ gives the scale of AM splitting of magnons, and $\left<ij\right>_{x,y}$ denotes the nearest neighbors in the $x,y$ directions, respectively.
Under the HP transformation and a Fourier transformation, the corresponding momentum-space magnon Hamiltonian amounts to
	\begin{equation}\begin{aligned}\label{k-Hamiltonian}
    \mathcal{H}(\boldsymbol{k})  &=  \left(A_1+A_2 \right) \sigma_0 \tau_0 + \left(A_1-A_2 \right) \sigma_0 \tau_3 \\&+ B_1 \sigma_1 \tau_1 + B_2 \sigma_0 \tau_1 + B_3\sigma_1\tau_0.
	\end{aligned}\end{equation}
	In the above, the model parameters read
    \begin{equation}
        \begin{aligned}
            A_{1,2} =&\,S\big[J_{1,2} \left(1-\cos{k_x}\right) +J_{2,1}\left(1-\cos{k_y}\right)   \\
            &+K \left(3\cos^2\theta-1\right)/4-\cos{2\theta }J_{\text{AFM}}/2 \\&+ g\mu_{\text{B}} H_0 \sin{\theta}/2 \big], \\
            B_{1}=&\, S J_{\text{AFM}}\left( 1+\cos{2\theta}\right)/2, \\
           B_{2}  =& - SJ_{\text{AFM}}\left( 1-\cos{2\theta}\right)/2, \\
           B_3 =& - S K \sin^2\theta/4, 
        \end{aligned}
    \end{equation}
    where $\theta = \arctan\left[g\mu_{\text{B}} S H_0 /\left( -2 J_{\text{AFM}} + K\right) \right] $ is the \mbox{canting} angle induced by the external magnetic field $H_0$, and $\sigma_{0,1,2,3}$ and $\tau_{0,1,2,3}$ denote the Pauli matrices operating on sublattice and wave vector degree of freedom, \mbox{respectively}.

\sect{Appendix B: Relation between magnon chirality and quantum metric} In the following, we provide further technical details on the unique quantum geometry of AM magnons retrieved and probed by light in this work. The magnonic metric $g^{(\text{m})}_{ab}$ considered in the main text reads in terms of magnon band eigenvectors $\ket{u_i} \equiv [U]_i$,
\begin{equation}\label{metric}
    g_{ab}^{(\text{m})} = \text{Re}  \frac{([\bar{U}]_{1} \tau_3 \partial_{k_a} \mathcal{H}(\boldsymbol{k}) [U]_{2})([\bar{U}]_{2} \tau_3 \partial_{k_b} \mathcal{H}(\boldsymbol{k}) [{U}]_{1})}{(\hbar\omega_1 -  \hbar\omega_2)^2},
\end{equation}
which is consistent with Refs.~\cite{owerre2017noncollinear, Bostrom2023}. We emphasize that the quantum geometry of magnons driven by the considered BdG Hamiltonians is fundamentally a symplectic quantum geometry~\cite{owerre2017noncollinear,Chaudhary2021, Bostrom2023, Tesfaye2025quantum}. We~now detail how the nontrivial metric texture [Fig.~\ref{LMC}(a)], which reflects the chirality $(p_j)$ alternation, and its coincidence with the $L^{(2)}_{ab}$ pattern [Fig.~\ref{LMC}(b)] identified in the main text, arise. The enhancement of diagonal terms $g^{(\text{m})}_{aa}$ [Fig.~\ref{LMC}(a)] occurs under a local $\boldsymbol{k}$-space condition: $|([\bar{U}]_{1} \tau_3 \partial_{k_a} \mathcal{H}(\boldsymbol{k}) [U]_{2})|^2 \gg 1$. These matrix elements, equivalently translate to a local enhancement of matrix elements $|\bar{U}^{\dagger}U_a|$ occurring in the LMC matrix elements, i.e., $L_{a}^{\left(1\right)}$ and $L_{ab}^{\left(2\right)}$, as defined in the main text, and consistently with the identifications of Ref.~\cite{Bostrom2023}. 

Crucially, the matrix elements $|\bar{U}^{\dagger}U_a|$ closely reflect the magnon chirality features.
On direct differentiation, the gradient $\partial_a p_j$, introduced by the unique alternation of AM magnonic chirality $(p_j)$, by definition consists of the terms $(U^{-1}_{ij})^* \partial_a U^{-1}_{ij}$.
The gradients of $U^{-1}_{ij}$ inverses in the individual magnon bands~$i$, within the $(U^{-1}_{ij})^* \partial_a U^{-1}_{ij}$ terms, translate into the enhancements in the matrix elements $|\bar{U}^{\dagger}U_a|$. 
A general inequality relating the quantum metric to the chirality alternation exists for collinear AM and AFM models in the limit $H_0\rightarrow 0$:
\begin{equation}
    \left|g_{aa}^{(\mathrm{m})}\right|
    \ge
    \left|\frac{\partial_a p_i}{2\Delta_{\boldsymbol{k}}}\right|^2,
\end{equation}
with $\Delta_{\boldsymbol{k}}$ arising through Bogoliubov transformation, as derived in SM~\cite{SM}. 
This inequality saturates for large-wave-vector magnons, which indicates that the enhancement of the quantum metric becomes a~direct signature of alternating magnon chirality at the BZ boundary.
To visualize these relations quantitatively, we presented the $\boldsymbol{k}$-space resolved local patterns in Figs.~\ref{geometric}(a) and~\ref{geometric}(b) against the textures in Figs.~\ref{LMC}(a) and~\ref{LMC}(b), and in Sec.~II of SM~\cite{SM}.

\end{document}


\title{{\textsc{SUPPLEMENTAL MATERIAL}} \\
    Quantum Geometry of Altermagnetic Magnons Probed by Light}

\newcommand{\TCM}{{TCM Group, Cavendish Laboratory, University of Cambridge, J.\,J.\,Thomson Avenue, Cambridge CB3 0HE, United Kingdom}}
\newcommand{\MCR}{Department of Physics and Astronomy, University of Manchester, Oxford Road, Manchester M13 9PL, United Kingdom}


\author{Rundong Yuan}
\email{ry306@cam.ac.uk}
\affiliation{\TCM}

\author{Wojciech J. Jankowski}
\email{wjj25@cam.ac.uk}
\affiliation{\TCM}

\author{Ka Shen}
\affiliation{The Center for Advanced Quantum Studies, School of Physics and Astronomy, and Institute for Advanced Study, Beijing Normal University, Beijing 100875, China}
\affiliation{Key Laboratory of Multiscale Spin Physics, Ministry of Education, Beijing Normal University, Beijing 100875, China}

\author{Robert-Jan Slager}
\email{rjs269@cam.ac.uk}
\affiliation{\MCR}
\affiliation{\TCM}

\date{\today}

\maketitle

\tableofcontents 

\newpage

\section{Light-magnon coupling}
\subsection{Perturbative Hamiltonian up to the fourth order}
The general light-magnon coupling up to the fourth order in the electromagnetic fields reads,
	\begin{equation}\label{generalLMC}
		\mathcal{H}({\boldsymbol{k}},\boldsymbol{A}) = \mathcal{H}(\boldsymbol{k}) + L^{\left(1\right)}_{a} A_a 
		+L^{\left(2\right)}_{ab}A_a A_b
		+L^{\left(3\right)}_{abc}A_a A_b A_c
		+L^{\left(4\right)}_{abcd}A_a A_b A_c A_d,
	\end{equation}	
     where $\boldsymbol{A} = \left(A_x, A_y, A_z\right)$ is the electromagnetic vector potential. $\boldsymbol{L}^{\left(n\right)}$ are the light-matter coupling (LMC) matrix elements~\cite{Topp2021, Bostrom2023}.
	Upon applying the normalization condition to Bogoliubov transformation we obtain an effective Schr\"odinger equation $\mathcal{H}(\boldsymbol{k})U = \bar{U}\Omega({\boldsymbol{k}})$. 
	On differentiating both sides and multiplying by $U^{\dagger}$ on the left, we obtain: 
	\begin{equation}\label{first-order}
		U^{\dagger}\mathcal{H}_a(\boldsymbol{k}) U = \Omega_a(\boldsymbol{k}) + U^{\dagger}\bar{U}_a \Omega(\boldsymbol{k}) - \Omega(\boldsymbol{k}) \bar{U}^{\dagger} U_a,
	\end{equation}
	where we use a matrix with subscript $a$ to represent that the matrix is differentiated by $\partial / \partial k_a$, and $a$ runs over all directions $x,y,z$.
	Eq.~\eqref{first-order} gives the first-order LMC coefficient as $L_a^{\left(1\right) } = -	U^{\dagger}\mathcal{H}^{\boldsymbol{k}}_a U$. 
	Similarly, we can differentiate the effective Schr\"odinger equation two, three, and four times to deduce $L_{ab}^{\left(2\right) }$, $L_{abc}^{\left(3\right) }$, and $L_{abcd}^{\left(4\right) }$.
	Consistently, the subscripts $b,c,d$ also run over all $x,y,z$.
	The LMC coefficients up to the fourth order are compactly retrieved as follows:
\begin{subequations}
    \begin{align}
L_a^{(1)} &= -\Omega_a(\bm{k}) - U^\dagger \bar{U}_a \Omega(\bm{k}) + \Omega(\bm{k}) \bar{U}_a^\dagger U, \\[2ex]
L_{ab}^{(2)} &= L_a^{(1)} \bar{U}^\dagger U_b - \frac{1}{2}\Omega \bar{U}^\dagger U_{ab} + \frac{1}{2} U^\dagger \bar{U}_{ab}\Omega + U^\dagger \bar{U}_a \Omega_b + \frac{1}{2}\Omega_{ab} + (ab \to \{ab\}),\label{eq:3b} \\[2ex]  
L_{abc}^{(3)} &= -\frac{1}{6} U^\dagger \bar{U}_{abc}\Omega - \frac{1}{2} U^\dagger \bar{U}_{ab}\Omega_c - \frac{1}{2} U^\dagger \bar{U}_a \Omega_{bc} - \frac{1}{6}\Omega_{abc} + \frac{1}{2} L_{ab}^{(2)} \bar{U}^\dagger U_c - \frac{1}{2} L_a^{(1)} \bar{U}^\dagger U_{bc} \nonumber \\
&\quad + \frac{1}{6}\Omega \bar{U}^\dagger U_{abc} + (abc \to \{abc\}), \\[2ex]
L_{abcd}^{(4)} &= \frac{1}{6} L_{abc}^{(3)} \bar{U}^\dagger U_d - \frac{1}{4} L_{ab}^{(2)} \bar{U}^\dagger U_{cd} + \frac{1}{6} L_a^{(1)} \bar{U}^\dagger U_{bcd} - \frac{1}{24}\Omega \bar{U}^\dagger U_{abcd} + \frac{1}{24} U^\dagger \bar{U}_{abcd}\Omega \nonumber \\
&\quad + \frac{1}{6} U^\dagger \bar{U}_{abc}\Omega_d + \frac{1}{4} U^\dagger \bar{U}_{ab}\Omega_{cd} + \frac{1}{6} U^\dagger \bar{U}_a \Omega_{bcd} + \frac{1}{24}\Omega_{abcd} + (abcd \to \{abcd\}).
\end{align}
\end{subequations}
where $+ \left(abcd \rightarrow \left\{abcd \right\} \right)$, etc., indicate that the equation is summing over all permutations. Below, we show the momentum-space profile of $L_{abcd}^{\left(4\right)}$ in the $d$-wave altermagnet model studied in the main text, against $L_{abcd}^{\left(4\right)}$ realized in an antiferromagnetic phase of the model ($\Delta J = 0$). 
We can also transform the LMC terms to quantum-metric related ones, using the relation of Ref~\cite{owerre2017noncollinear},
\begin{equation}\label{metric_sm}
    g_{ab}^{(\text{m})} = \text{Re}\left[ \bar{U}^\dagger_a U_b\right].
\end{equation}
The first term in Eq.~\eqref{eq:3b} contains the metric $g_{ab}^{(\text{m})}$, which further simplifies the form of LMC shown in the main text.

 \begin{figure}
        \centering
        \includegraphics[width=0.69\columnwidth]{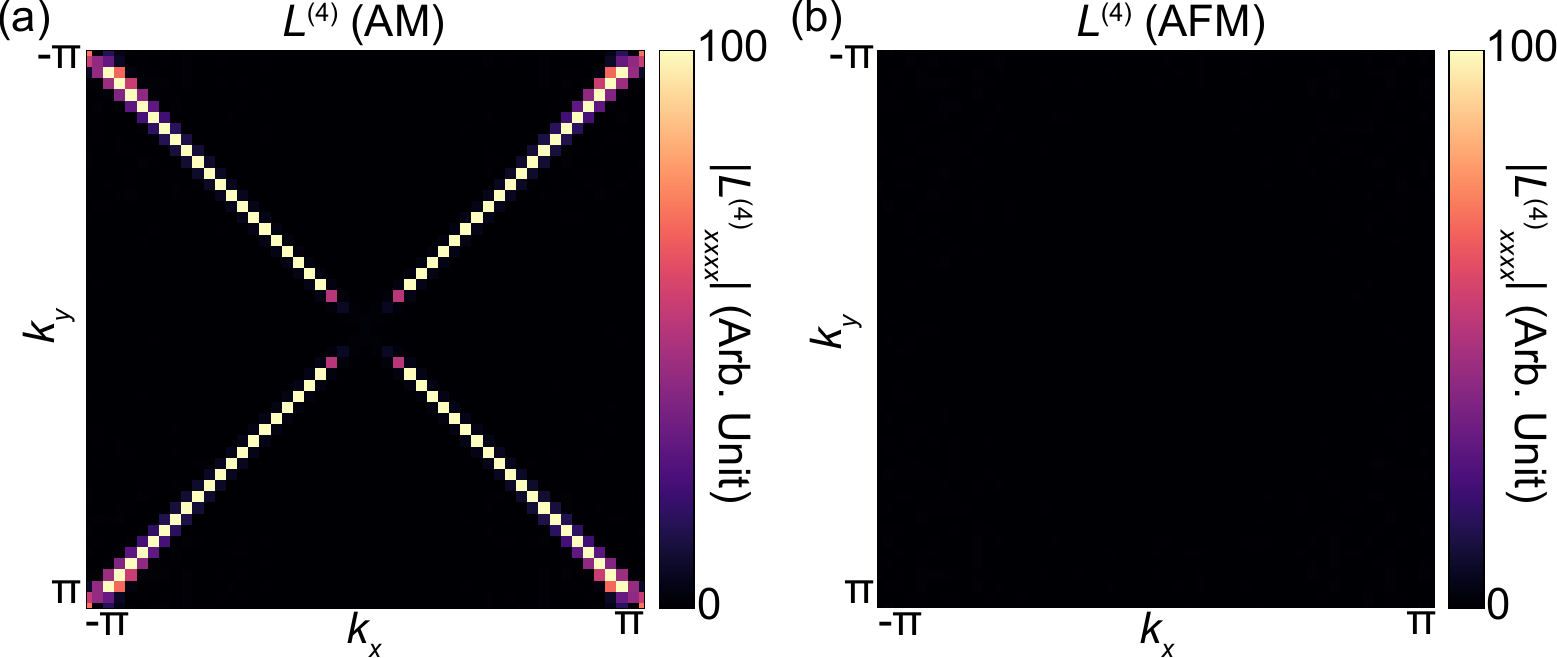}
        \caption{Fourth-order light-magnon coupling matrix elements, $L_{abcd}^{\left(4\right)}$, in (a) an altermagnetic (AM) model studied in the main text, and (b) the corresponding antiferromagnetic (AFM) model. Although, the associated fourth-order responses are negligible compared to the second-order nonlinear coupling $L_{ab}^{\left(2\right)}$ central to the main text, the AFM notably realizes orders-of-magnitude smaller fourth-order nonlinearity, as compared to the AM phase.}
        \label{L4}
    \end{figure}

\newpage

\subsection{Fleury-Loudon theory of light-magnon coupling}
In this work, to determine the bicircular Raman responses, we consider light-magnon coupling within Fleury-Loudon theory. The Fleury-Loudon type of light-magnon coupling involves two magnons with opposite wave vectors, scattered by the light, which gives the scattering cross-section $P(\omega_i)$ as follows~\cite{fleury1968scattering, Bostrom2023, liu2026light}
\begin{equation}
    P(\omega_i) \propto \left| \bra{u_i^{\text{(m)}}}\mathcal{H}_{\text{R}}\ket{u_{\bar{i}}^{\text{(m)}}}\right|^2,
\end{equation}
where $P(\omega_i)$ captures the response from the magnons with energy $\hbar\omega_i$, $ \mathcal{H}_{\text{R}}$ is a perturbative Raman Hamiltonian, and $i(j)$ the band index of magnon states $|u_{i}^{\text{(m)}} \rangle$, and $\bar{i}$ denotes an opposite wave vector within the magnon band~$i$.
It should be emphasized that the $\langle u_{i}^{\text{(m)}} |$ and $|u_{\bar{i}}^{\text{(m)}} \rangle$ here denote the magnon eigenvector rather than electron wavefunctions.
The Raman Hamiltonian is proportional to the dipolar coupling between light and ions.
In the approximate limit of $\omega_{\text{in}}\approx\omega_{\text{sc}}$, where $\omega_{\text{in}}$ and $\omega_{\text{sc}}$ are the frequencies of the incident and scattered light, respectively, we employ the standard Fleury-Loudon LMC Hamiltonian based on the Hubbard model as (see also Ref.~\cite{liu2026light}):
\begin{equation}
\mathcal{H}^{\mathrm{FL}}(\boldsymbol{k}, \boldsymbol{A})
=
\frac{e^2}{\hbar^2}
\frac{1}{U - \omega_{\mathrm{in}}}
\left( \boldsymbol{A}_{\mathrm{sc}}^{*} \cdot \boldsymbol{d}_{ij} \right)
\left( \boldsymbol{A}_{\mathrm{in}} \cdot \boldsymbol{d}_{ij} \right)
\mathcal{H}(\boldsymbol{k}),
\end{equation}
where $U$ denotes the Hubbard energy, $\boldsymbol{A}_{\text{sc(in)}}$ are the vector potentials of scattered (incident) light, and $\boldsymbol{d}_{ij}$ is the position vector from spin $i$ to spin $j$~\cite{fleury1968scattering}. 
This mechanism indicates that the light-magnon scattering process is a~two-magnon one.
The vector potential follows a long-wavelength approximation $\boldsymbol{A}\cdot\boldsymbol{d}_{ij} \ll 1$, which further allows to approximate $ \boldsymbol{d}_{ij} \rightarrow-i\partial_{\boldsymbol{k}}$~\cite{ahn2022riemannian}. 
Thus, upon action of the displacement operators as derivatives in the reciprocal space, we then have the perturbative Hamiltonian that reads
\begin{equation}
    \mathcal{H}_{\text{R}}^{(2)}\propto \sum_{a,b}  L_{ab} A_{\text{in},a} A_{\text{sc},b},
\end{equation}
where $a,b = x,y,z$ are the spatial direction indices, and the LMC terms are defined as $L_{ab} = \partial_{k_a}\partial_{k_b}\mathcal{H}(\boldsymbol{k})$.
A detailed formal procedure deriving the perturbative Hamiltonian $\mathcal{H}_{\text{R}}^{(2)}$ from Fleury-Loudon theory can also be found in Ref.~\cite{liu2026light}.

\newpage
	
\section{Comparison of models with different parameter choices}

\subsection{Altermagnetic vs. antiferromagnetic models}

    \begin{figure}[t!]
        \centering
        \includegraphics[width=0.96\columnwidth]{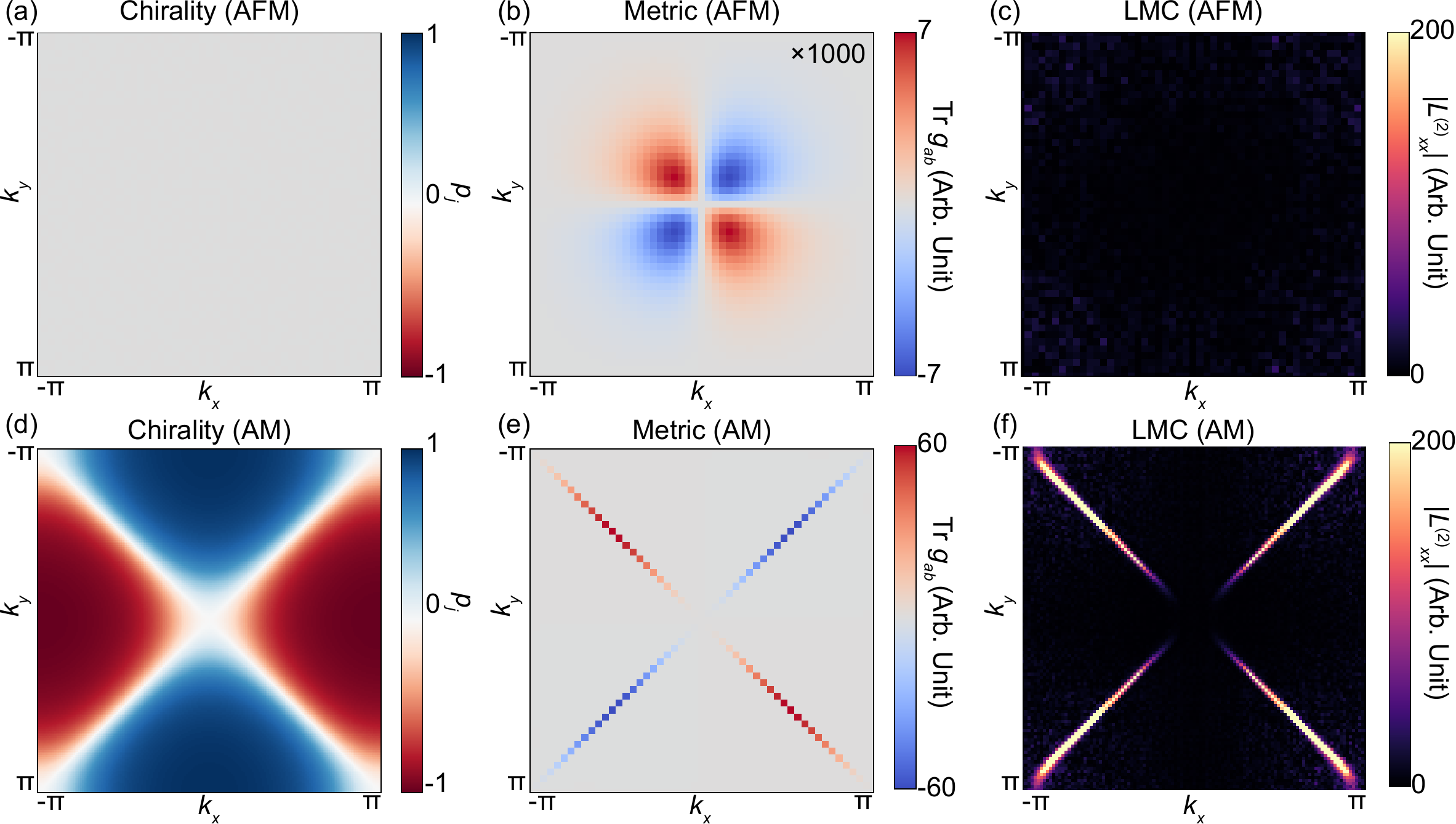}
        \caption{Comparison of the lowest magnon band chiralities (a,d), magnonic quantum metrics (b,e), and light-magnon couplings (c,f) in an antiferromagnetic (AFM) model (a)--(c), against the altermagnetic (AM) model (d)--(f) studied in the main text.}
        \label{AFM}
    \end{figure} 

In this Section we compare our main findings in altermagnets against the conventional antiferromagnets.
The parameterization of our model is the same as in the main text, with a different altermagnetic splitting: $\Delta J = 0$. We also consistently apply a magnetic field along the $x$ direction.
In Fig.~S\ref{AFM}(a), we show the chirality of magnons in antiferromagnets with the same color bar as for the altermagnet in the main text. 
In the considered model, we find that the magnons with left- and right- handed chirality are degenerate over the full Brillouin zone (BZ) in the antiferromagnetic state. The degeneracy is expected, since the magnetic field is perpendicular to the N\'eel vector.
Hence, there is no altering chirality in the antiferromagnetic model, unlike in the corresponding altermagnetic case.
We then calculate the magnonic quantum metric of the system, and find that the metric, and its trace, are much smaller than in the altermagnetic setting, as shown in Fig.~S\ref{AFM}(b). We note that in the main text we retrieve a vast range of values ($\pm 60$), as captured by the color bar, while in Fig.~S\ref{AFM}(b), a significantly smaller range of $\pm 0.007$ is found within an antiferromagnetic phase.
The quantum metric result is compatible with the claim of our main text that without altering the chirality across BZ, there is no nontrivial metric texture in the system. We also calculate the $\boldsymbol{k}$-resolved nonlinear LMC terms which is shown in~Fig.~S\ref{AFM}(c).
As represented using the same color bar as in the main text, we find three orders of magnitude smaller response encoded in the LMC within the antiferromagnetic model, which also verifies our main claim. Furthermore, for a direct comparison, we include the analogous results for the altermagnet studied in the main text [Fig.~S\ref{AFM}(d)--Fig.~S\ref{AFM}(f)], to manifestly show these major distinctions between the antiferromagnetic and altermagnetic magnon states.

\begin{figure}[t!]
        \centering
        \includegraphics[width=\columnwidth]{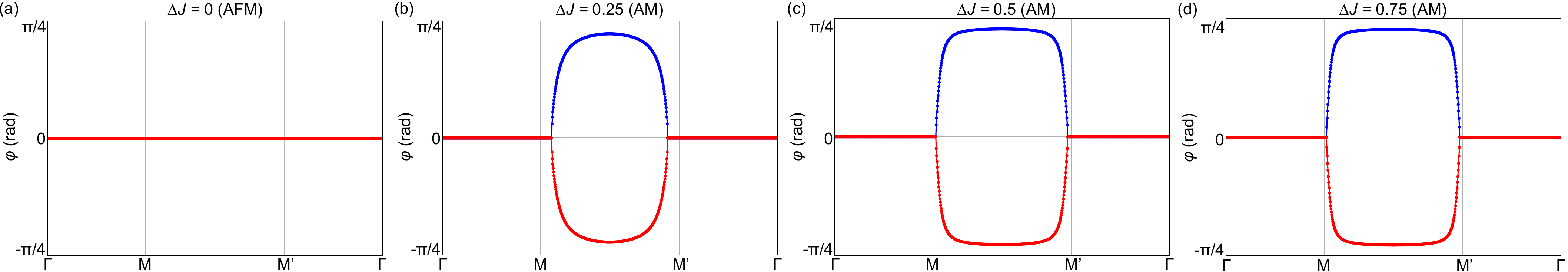}
        \caption{Comparison of the magnonic Wilson loops in (a) antiferromagnetic (AFM) phase ($\Delta J = 0$), against the (b)--(d) altermagnetic (AM) magnons ($\Delta J \neq 0$).}
        \label{wilson}
\end{figure} 

Finally, in Fig.~S\ref{wilson}, we directly demonstrate how the geometric phases captured by the magnonic Wilson loop compare between the antiferromagnetic ($\Delta J = 0$) [Fig.~S\ref{wilson}(a)] and altermagnetic ($\Delta J \neq 0$) [Fig.~S\ref{wilson}(b)--Fig.~S\ref{wilson}(d)] model realizations. We confirm that, unlike in the altermagnetic magnon bands, the antiferromagnets realize no nontrivial features in magnonic geometry. Namely, we find that the magnonic Wilson loop eigenvalues are trivial ($\phi=0$) in the antiferromagnet, consistently with the chirality arguments provided in the main text. 
\newpage

\subsection{Collinear vs. canted altermagnets}
\begin{figure}[t!]
    \centering
    \includegraphics[width=\linewidth]{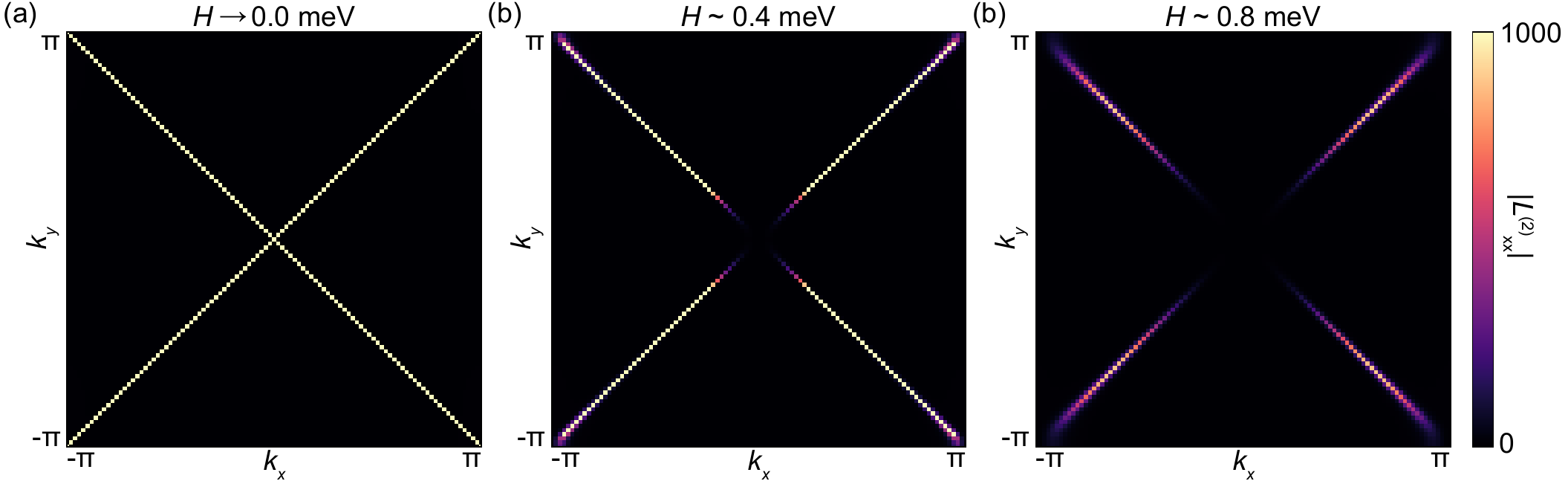}
    \caption{The evolution of LMC term as a function of the external field $H$, where $H$ matches the energy of (a) 0.0001~meV, (b) 0.4 meV, and (c) 0.8 meV.}
    \label{L22-H}
\end{figure}
To study the effects of canting field on light-magnon coupling, we further calculated the first diagonal term of the LMC coupling matrix, as a function of the canting external magnetic field $H$.
As the external field increases, the LMC strength is found to be eliminated, where one can find the difference between LMC with external field matches the energy scale of (a) 0 meV, (b) 0.4 meV, and (c) 0.8 meV. 
Although when $H_0=0$ there is a degeneracy of magnonic bands, we can still deduce such effects with a finite but small field. 
In the calculation, we apply an external field corresponding to an energy of 0.0001 meV.

\newpage

\section{Further analytical details}

\subsection{Relation between magnonic quantum metric and alternating magnon chirality}

In the following, we derive a bound on the magnonic quantum metric components $g^{(m)}_{aa}$, from the chirality $(p_j)$ alternation captured by the momentum-space derivatives $\partial_a p_j$. As defined in the main text, we consider the magnonic eigenvector matrix $V\equiv U^{-1}$, which satisfies the paraunitary normalization condition
\begin{equation}
V\tau_3 V^\dagger=\tau_3.
\end{equation}
The quantum metric matrix can be rewritten as the real part of the geometric tensor in the inverse-space basis, which reads
\begin{equation}
g_{ab}^{(\text{m})}=\text{Re}\left[V_{a}\bar{V}_b^\dagger\right],
\end{equation}
with $V_a\equiv\partial_{k_a}V$ and $\bar{V}_b^\dagger \equiv \partial_{k_b} \bar{V}^\dagger$. 
For the $i$-th row of the matrix $V$, we decompose the components as follows
\begin{equation}
[V]_i\equiv v_i=(v_{i,+},v_{i,-}),
\end{equation}
where $v_{i,+}$ and $v_{i,-}$ denote the positive- and negative- wavevector sectors of the $\tau_3$ metric, that is, $v_{i,+} =(v_{i1}, v_{i2})$, and $v_{i,-}=(v_{i3}, v_{i4})$.
Here, we consider the limit of a vanishingly small external field, $H_0 \rightarrow 0^+$, in which the altermagnetic magnon eigenmodes are lifted, but not significantly modified by the external field, such that the eigenmodes can be expressed as a linear combination of the zero-field wavefunctions (i.e. modes at $H_0 = 0$).
In this limit, we obtain the Bogoliubov transformation that yields \cite{Shen2020magnon,jin2025strong}
\begin{equation}
\|v_{i,+}\|=\sqrt{\frac{\Delta_{\boldsymbol{k}}+1}{2}},
\qquad
\|v_{i,-}\|=\sqrt{\frac{\Delta_{\boldsymbol{k}}-1}{2}},
\end{equation}
where $\Delta_{\boldsymbol{k}}
=\sqrt{
\frac{(A_1+A_2)^2}{(A_1+A_2)^2-B^2}
}$. 
We note that this solution is independent of the specific expressions of the parameters, and therefore holds for any collinear AM/AFM.
Since the diagonal terms of the quantum metric $\left[g_{aa}^{(\text{m})}\right]_{ii}$ measures the difference of derivative weights between the $+\boldsymbol{k}$ and $-\boldsymbol{k}$ components, i.e.,
\begin{equation}
\left[g_{aa}^{\text{(m)}}\right]_{ii}
=
\|\partial_a v_{i,+}\|^2
-
\|\partial_a v_{i,-}\|^2,
\end{equation}
we obtain a simple relation, \begin{equation}\label{deltak}
    |\partial_a\Delta_{\boldsymbol{k}}|^2
=
4(\Delta_{\boldsymbol{k}}^2-1)\left|\left[g_{aa}^{\text{(m)}}\right]_{ii}\right|
.
\end{equation}
We can then connect this to the chirality of magnonic band with index $i$, which can be rewritten as
\begin{equation}
p_i=v_i C v_i^\dagger,
\qquad
C=\mathrm{diag}(1,-1,-1,1),
\end{equation}
derivatives of which read,
\begin{equation}
\partial_a p_i
=
2\mathrm{Re}\!\left[
(\partial_a v_i)C v_i^\dagger
\right].
\end{equation}
Physically, $\partial_a p_i$ is the projection of the eigenvector change  $\partial_a v_i$ onto the direction $C v_i^\dagger$. 
Hence, it cannot exceed the total size of the eigenvector change. 
Such relation is captured by the Cauchy--Schwarz inequality,
\begin{equation}
|\partial_a p_i|^2
\le
4\|v_i\|^2\|\partial_a v_i\|^2.
\end{equation}
Using
\begin{equation}
\|v_i\|^2=\Delta_{\boldsymbol{k}}, \qquad
\text{and} \qquad
\|\partial_a v_i\|^2
=
\|\partial_a v_{i,+}\|^2+\|\partial_a v_{i,-}\|^2
=
\frac{\Delta_{\boldsymbol{k}}}
{4(\Delta_{\boldsymbol{k}}^2-1)}
|\partial_a\Delta_{\boldsymbol{k}}|^2,
\end{equation}
we find
\begin{equation}
|\partial_a p_i|^2
\le
\frac{\Delta_{\boldsymbol{k}}^2}
{\Delta_{\boldsymbol{k}}^2-1}
|\partial_a\Delta_{\boldsymbol{k}}|^2.
\end{equation}
Finally, substituting the relation obtained before in Eq.~\eqref{deltak}, we can deduce the absolute value bound
\begin{equation}\label{eq:CS}
\boxed{
\left|\left[g_{aa}^{\text{(m)}}\right]_{ii}\right|
\ge
\frac{\left|\partial_a p_i\right|^2}
{4\Delta_{\boldsymbol{k}}^2}
}.
\end{equation}
The inequality Eq.~(\ref{eq:CS}) is saturated when $\Delta_{\bm{k}} \to 1^+$ (i.e. $|A_1+A_2| \gg |B|$), where the quantum metric is fully determined by the derivative of the magnon chirality. 
In our model, we have $A_1+A_2 \sim  B$ near the $\Gamma$ point, and meanwhile $A_1+A_2 > B$ towards the Brillouin zone boundary. 
Since the two-magnon Raman scattering is dominated by magnons with finite wavevectors, the Raman response serves as a direct probe of the quantum metric, when the large quantum metric is determined, and hence guaranteed, by the saturation of Eq.~(\ref{eq:CS}).

\subsection{Phase selection of light-magnon coupling terms}
In the following, we provide further analytical details on the bicircular Raman response underpinned by the outlined LMC perturbations, as retrieved numerically in the main text. We further elucidate the phase delay ($\alpha$) dependence of the optical response. On inserting the bicircular light form introduced in the main text to the general LMC Hamiltonian, Eq.~\eqref{generalLMC}, we have,
\\
\begin{equation}
    \mathcal{H}(\boldsymbol{k},\boldsymbol{A}) = \mathcal{H}(\boldsymbol{k}) + L^{\left(1\right)}_{a} (\boldsymbol{A}_\text{L} e^{\text{i} ( \omega_\text{in} t + \alpha)} + \boldsymbol{A}_\text{R} e^{-\text{i} \omega_\text{in} t})_a 
		+L^{\left(2\right)}_{ab} (\boldsymbol{A}_\text{L} e^{\text{i} ( \omega_\text{in} t + \alpha)} + \boldsymbol{A}_\text{R} e^{-\text{i} \omega_\text{in} t})_a (\boldsymbol{A}_\text{L} e^{\text{i} ( \omega_\text{in} t + \alpha)} + \boldsymbol{A}_\text{R} e^{-\text{i} \omega_\text{in} t})_b.
\end{equation}
\\
Here we truncated the LMC at the second order in optical fields, as justified in the previous Sections. On setting equal strengths ($A_0$) for left- and right- circularly-polarized components $\boldsymbol{A}_\text{L/R} = A_0 \hat{\boldsymbol{e}}_\pm \equiv A_0\big(\frac{\hat{\boldsymbol{x}}\pm \text{i}\hat{\boldsymbol{y}}}{\sqrt{2}}\big)$, with $\hat{\boldsymbol{x}}, \hat{\boldsymbol{y}}$ the unit polarization vectors along $x$ and $y$ directions, we further simplify the considered bicircular coupling to,
\\
\begin{equation}
\begin{aligned}
    \mathcal{H}(\boldsymbol{k},\boldsymbol{A}) &= \mathcal{H}(\boldsymbol{k}) + \frac{|A_0|}{\sqrt{2}} L^{\left(1\right)}_{a} [(\hat{\boldsymbol{x}} + \text{i}\hat{\boldsymbol{y}})  e^{\text{i} ( \omega_\text{in} t + \alpha)} + (\hat{\boldsymbol{x}} - \text{i}\hat{\boldsymbol{y}})  e^{-\text{i} \omega_\text{in} t}]_a 
		\\&+\frac{|A_0|^2}{2} L^{\left(2\right)}_{ab} [(\hat{\boldsymbol{x}} + \text{i}\hat{\boldsymbol{y}}) e^{\text{i} ( \omega_\text{in} t + \alpha)} +  (\hat{\boldsymbol{x}} - \text{i}\hat{\boldsymbol{y}}) e^{-\text{i} \omega_\text{in} t}]_a [(\hat{\boldsymbol{x}} + \text{i}\hat{\boldsymbol{y}}) e^{\text{i} ( \omega_\text{in} t + \alpha)} +  (\hat{\boldsymbol{x}} - \text{i}\hat{\boldsymbol{y}}) e^{-\text{i} \omega_\text{in} t}]_b,
\end{aligned}
\end{equation}
\\
\begin{equation}
\begin{aligned}
    \mathcal{H}(\boldsymbol{k},\boldsymbol{A}) &= \mathcal{H}(\boldsymbol{k}) + \sqrt{2} |A_0| e^{\text{i}\alpha/2} L^{\left(1\right)}_{a}[ \hat{\boldsymbol{x}} \cos ( \omega_\text{in} t + \alpha/2)-\hat{\boldsymbol{y}} \sin ( \omega_\text{in} t + \alpha/2)]_a 
		\\&+2|A_0|^2 e^{\text{i}\alpha} L^{\left(2\right)}_{ab} [\hat{\boldsymbol{x}} \cos ( \omega_\text{in} t + \alpha/2)-\hat{\boldsymbol{y}} \sin ( \omega_\text{in} t + \alpha/2)]_a [\hat{\boldsymbol{x}} \cos ( \omega_\text{in} t + \alpha/2)-\hat{\boldsymbol{y}} \sin ( \omega_\text{in} t + \alpha/2)]_b.
\end{aligned}
\end{equation}
\\
The choice of individual indices $a,b=x,y$ in LMC selects distinct $\alpha$-dependent components: $\hat{\boldsymbol{x}} \cos ( \omega_\text{in} t + \alpha/2)$ and $\hat{\boldsymbol{y}}  \sin( \omega_\text{in} t + \alpha/2)$. For $ L^{\left(2\right)}_{xx}$ component shown in the main text, the strictly nonnegative term: ${|A_0|^2\cos^2 ( \omega_\text{in} t + \alpha/2)}$ competes with: $|A_0|\cos ( \omega_\text{in} t + \alpha/2)$, for the $L^{\left(1\right)}_{x}$ coupling. Notably, the $L^{\left(1\right)}_{x}$ and $L^{\left(2\right)}_{xx}$ terms have different phase-dependences in the corresponding amplitudes related to the power absorption, independently from the time-averaging of the oscillatory terms. Averaging the cross-sections over the range of bicircular phase delay parameter $\alpha$ allows to distinguish the magnitudes of $L^{\left(1\right)}_{x}$ and $L^{\left(2\right)}_{xx}$ related to the magnonic metric $g^{(\text{m})}_{ab}$, as pointed out in the main text. More precisely, inserting the LMC in the Fermi's golden rule expression~\cite{Topp2021} for Raman scattering, the matrix elements of the above bicircular LMC, combined with their conjugates, introduce distinct phase-shift dependences on $\alpha$ for $L^{\left(1\right)}_{a}$ and $L^{\left(2\right)}_{ab}$ contributions. Hence, the dependences of the bicircular Raman scattering cross-section on on $\alpha$ allows to specifically target~$L^{\left(2\right)}_{ab}$, which reflects the magnonic quantum metric tensor $g^{(\text{m})}_{ab}$~\cite{Topp2021}, as described in the main text.

\subsection{Circular vs. linear basis for light-magnon coupling}

For further comparison, we phrase the perturbative LMC in the circular polarization basis of light.
We begin with the transformation of the field components $A_x = (A_{\text{L}}+A_{\text{R}})/2$, and $A_y = (A_{\text{L}}-A_{\text{R}})/(2\text{i})$. 
We can then transform the LMC terms into 
\begin{equation}
    \mathcal{H}(\boldsymbol{A})=\tilde{L}^{(1)}_{p}A_p + \tilde{L}^{(2)}_{pq}A_p A_q+O(A^3),
\end{equation}
where the $\tilde{\bf{L}}^{(n)}$ denotes the LMC matrix components in the circularly-polarized basis, and the subscripts $p,q$ denote the left-handed (LH) and right-handed (RH) circular light polarizations. We use $p,q = \pm$ to denote the unit polarization vectors of LH/RH circularly-polarized light: $\hat{\boldsymbol{e}}_{\pm}=\frac{\boldsymbol{\hat{x}}\pm \text{i} \boldsymbol{\hat{y}}}{\sqrt{2}}$. Transforming the LMC matrix elements from linear to circular polarization basis obtains,
\begin{subequations}
    \begin{align}
        L_{++} &= \frac{1}{4}(L_{xx}-L_{yy})+\frac{1}{4\text{i}}(L_{yx}+L_{xy}), \\
        L_{--} &= \frac{1}{4}(L_{xx}-L_{yy})-\frac{1}{4\text{i}}(L_{yx}+L_{xy}), \\
        L_{+-} &= L_{-+} = \frac{1}{4}(L_{xx}+L_{yy}), 
    \end{align}
\end{subequations}
which clearly shows that the responses to different polarizations of light are different.
We can then rewrite the BCL Raman response due to the light-magnon interactions in terms of the Raman tensor $R_{ab}$. In different basis representations, the Raman tensor reads, 
\begin{subequations}
    \begin{align}
            R_{ab}&=\begin{pmatrix}
            L_{xx}^{(2)},L_{xy}^{(2)}\\
            L_{yx}^{(2)}, L_{yy}^{(2)}
        \end{pmatrix},\\
         \tilde{R}_{s_1 s_2}&=\begin{pmatrix}
                        L_{++}^{(2)},L_{+-}^{(2)}\\
                        L_{-+}^{(2)}, L_{--}^{(2)}
                        \end{pmatrix},
    \end{align}
\end{subequations}
where $s_{1,2}$ denote different circular light polarizations. The contraction of the Raman tensor with different polarization vectors obtains the $P_\text{BC}$ matrix elements: $|R_{ab} \hat{\boldsymbol{e}}_a \hat{\boldsymbol{e}}_b|^2$. Notably, the bicircular light spans a complete basis set of light mode polarizations, when $\eta = 1$.
However, for $\eta\neq1$, i.e., the RH and LH components of BCL are not at the same frequency, the bicircular basis is incomplete for describing the possible polarizations of light, since the RH and LH components are mutually incoherent in that case.
However, in our work, we consider a special condition with $\eta=1$, which can be employed as a complete basis.
If we set $\hat{\boldsymbol{e}}_{1}^{\text{BC}}=\frac{1}{\sqrt{2}}(\hat{\boldsymbol{e}}_+ + \text{i}\hat{\boldsymbol{e}}_-)$, $\hat{\boldsymbol{e}}_{2}^{\text{BC}}=\frac{1}{\sqrt{2}}(\hat{\boldsymbol{e}}_+ - \text{i}\hat{\boldsymbol{e}}_-)$, we can construct the complete basis in a one-to-one correspondence to the circular basis.

In the circular basis, we can then deduce the Raman cross-section with the incident and scattered bicircular polarizations of BCL, $\boldsymbol{A}_{\text{L}}e^{\text{i} (\Omega t+\alpha)}+\boldsymbol{A}_{\text{R}}e^{- \text{i} \Omega t}$, which we have
\begin{equation}
   P_{\text{BC}} \propto \left|U^\dagger\left(L^{(2)}_{++}A_{\text{L}}^2 e^{2\text{i}\alpha}+2L_{+-}^{(2)}A_{\text{L}}A_{\text{R}}e^{\text{i}\alpha} +L_{--}^{(2)}A_{\text{R}}^2\right)U\right|^2 \delta(\omega-\omega_{\text{sc}}+\omega_{\text{in}}),
\end{equation}
where a time averaging is applied. One finds that the Raman scattering is eliminated when $\alpha=\pi/2$, and when $A_{\text{L}}=A_{\text{R}}\equiv|A|$ condition is set.
The Raman scattering cross-section follows as $ P_{\text{BC}}^{\pi/2} \propto \left|U^\dagger|A|^2\left(-L^{(2)}_{++}+2\text{i}L_{+-}^{(2)}+L_{--}^{(2)}\right)U\right|^2 = 0 $, which corresponds to the blue dotted line of the main text Fig.~4(a), where no scattering is expected.

\bibliography{references}